\documentclass[11pt]{article}
\usepackage{moriond,epsfig}

\def\be{\begin{equation}}
\def\ee{\end{equation}}
\def\bea{\begin{eqnarray}}
\def\eea{\end{eqnarray}}

\def\gsim{\mathrel{\rlap{\lower4pt\hbox{\hskip1pt$\sim$}}
    \raise1pt\hbox{$>$}}}         
\newcommand{\ket}[1]{| {#1} \rangle}
\newcommand{\beq}{\begin{equation}}
\newcommand{\eeq}{\end{equation}}
\newcommand{\br}{{\bf r}}
\def\lsim{\mathrel{\rlap{\lower4pt\hbox{\hskip1pt$\sim$}}
  \raise1pt\hbox{$<$}}}         
\begin{document}
\vspace*{4cm}
\title{FULL OF CHARM NEUTRINO DIS}

\author{ R.~Fiore$^{1}$ and V.R.~Zoller$^{2}$}

\address{$^1$Dipartimento di Fisica,
Universit\`a     della Calabria\\
and\\
 INFN, Gruppo collegato di Cosenza, Italy\\
$^2$ITEP, Moscow 117218, Russia}

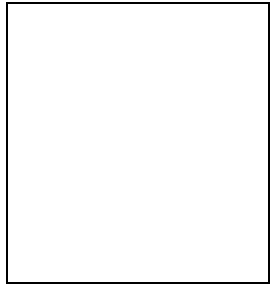
\maketitle\abstracts{
The color dipole analysis of  small-$(x, Q^2)$ neutrino DIS induced by 
the charmed-strange ($cs$) current
 reveals ordering of 
dipole sizes $m_c^{-2}<r^2<m_s^{-2}$ typical of the Double Leading Log 
Approximation (DLLA).
The DLLA resummation leads to the $cs$ component of the 
longitudinal structure function 
$F_L$ rising to small $x$ much faster than its light quark component.
 Based on the color dipole BFKL approach
 we report quantitative 
predictions for this effect in the kinematical range of the 
CCFR/NuTeV experiment. }

 We  report our  analysis of  the charged current (CC) 
non-conservation effects in  small-$x$  neutrino DIS.
We use  the color dipole (CD) basis  of high-energy QCD \cite{NZ91,M} 
and  quantify  the phenomenon of weak current non-conservation
in terms  of the 
light cone wave functions (LCWF) $\Psi^{cs}_{\lambda}$
 in the Fock  expansion of the  $W^+$-boson state with helicity $\lambda$,
\beq
\ket{W^+_{\lambda}}=\Psi^{cs}_{\lambda}\ket{c\bar s}
+\Psi^{ud}_{\lambda}\ket{u\bar d}+...
\label{eq:FOCK}
\eeq 
At small $Q^2\lsim m_c^2$
the strong un-equality of masses of the charmed and strange quarks 
manifests its effects and  the CD   analysis  reveals the
 ordering of dipole sizes
\beq 
(m_c^{2}+Q^2)^{-1}\lsim r^2 \ll m_s^{-2}
\label{eq:ORDER}
\eeq
typical of the Double Leading Log Approximation (DLLA) \cite{D}. 
 The  multiplication of $\log$'s 
like 
\beq
\alpha_S\log[(m_c^2+Q^2)/\mu^2_G)\log(1/x) 
\label{eq:LOGLOG}
\eeq
to
higher orders 
of perturbative QCD
ensures the dominance of the charmed-strange component, $F_L^{cs}$,
 of the longitudinal structure function (LSF)
\beq
F_L=F_L^{ud}+F_L^{cs}
\label{eq:csud}
\eeq 
in the kinematical domain  covered by  the CCFR/NuTeV experiment 
\cite{Fleming}.

In  the vacuum exchange dominated  region
of $x\lsim 0.01$ 
the contribution of excitation of open 
charm/strangeness to the longitudinal 
($\lambda=L$)
and transverse  ($\lambda=T$) structure functions
is given by 
\beq
F_{\lambda}(x,Q^{2})
={Q^2\over 4\pi^2\alpha_{W}}\int dz d^{2}{\bf{r}}
|\Psi_{\lambda}(z,{\bf{r}})|^{2} 
\sigma(x,r)\,,
\label{eq:FACTOR}
\eeq
where $\alpha_W=g^2/4\pi$ and the weak charge $g$ is
related to the Fermi coupling constant $G_F$,
${G_F/\sqrt{2}}={g^2/m^2_{W}}$.
$|\Psi_{\lambda}(z,{\bf{r}})|^2$  is the light 
cone density of 
$c\bar s$ dipoles of the size ${\bf r}$  with the $c$ quark 
carrying fraction $z$ of the $W^+$ light-cone momentum. In particular,
 $|\Psi_{L}|^2$ 
is the incoherent sum of   the vector $(V_L)$  and the  
axial-vector $(A_L)$ terms:
$
|\Psi_{L}|^2= |V_{L}|^2+ |A_{L}|^2$.
At $Q^2\gg m_c^2$ the $S$-wave component of  
$\ket{c\bar s}$ dominates \cite{Kolya92,FZ1},
\bea
|V_L|^2\sim |A_L|^2\propto
Q^2z^2(1-z)^2K_0^2(\varepsilon r).
\label{eq:RHOS1}
\eea
At $Q^2\lsim m_c^2$ the $P$-wave, that arises
 due to the current non-conservation, takes over,
\bea
|V_L|^2\sim|A_L|^2\propto
 {m_c^2\over Q^2}
\varepsilon^2 K^2_1(\varepsilon r).
\label{eq:RHOS2}
\eea
Here
$\varepsilon^2=z(1-z)Q^2+(1-z)m_c^2+zm_s^2$ controls the size of  $c\bar s$ 
dipole, $r^2\sim \varepsilon^{-2}$.  

At small $Q^2\lsim m_c^2$  
integrating  over  $z$ for $r^2$ from the region defined by 
the inequality (\ref{eq:ORDER})  yields \cite{FZCS}
\bea
\int dz |\Psi_{L}(z,\br)|^2\approx {\alpha_WN_c\over \pi^2}{m_c^2\over
 m_c^2+Q^2}{1\over Q^2r^4}
\label{eq:FLZ}
\eea
so that Eqs.(\ref{eq:FACTOR},\ref{eq:FLZ}) give rise to nested logarithmic 
integrals over dipole sizes.

In the CD  approach the BFKL-$\log(1/x)$ evolution \cite{BFKL} 
of  $\sigma(x,r)$
is described by the CD BFKL equation of Ref.\cite{NZZBFKL}. 
For qualitative estimates it suffices to use  the DLLA. 
In the Born approximation (2g-exchange) \cite{NZ91}   
\beq
\sigma(r)\approx {C_F\pi^2 }r^2\alpha_S(r^{-2})L(r^{-2}) .
\label{eq:2G}
\eeq
where 
\beq
L(k^2)={4\over \beta_0}
\log{\alpha_S(\mu_G^2)\over\alpha_S(k^2)}.
\label{eq:L}
\eeq
 and $\alpha_S(k^2)=4\pi/\beta_0\log(k^2/\Lambda^2)$ with
$\beta_0=11-2N_f/3$.

Perturbative gluons do not propagate to large distances and $\mu_G$
stands  for the inverse Debye screening radius, $\mu_G=1/R_c$.
The lattice QCD data suggest $R_c\approx 0.3$ fm \cite{MEGGIO}.
Because $R_c$    is small  compared to the typical range of strong 
interactions, the dipole cross section  evaluated with  the  decoupling of 
soft gluons, $k^2\lsim \mu_G^2$,
 would underestimate  the interaction strength for
 large color dipoles. In Ref.\cite{NPT,NSZZ,LANACH} this missing strength
 was modeled by 
a non-perturbative, soft correction $\sigma_{npt}(r)$ to the 
dipole cross section $\sigma(r)=\sigma_{pt}(r)+\sigma_{npt}(r).$ 
Here we concentrate on  the perturbative component, $\sigma_{pt}(r)$,
represented by Eq.(\ref{eq:2G}).

Then, at $Q^2\lsim m_c^2$ 
\beq
F_L^{cs}\sim 
{N_cC_F\over 4}{m_c^2\over {m_c^2+Q^2}}{1\over {2!}}L^2(m_c^2+Q^2).
\label{eq:FL2G}
\eeq
There is also a  contribution to 
$F_L^{cs}$  from the region $0<r^2<(m_c^2+Q^2)^{-1}$
\bea  
F_L^{cs}\sim 
{N_cC_F\over 4}{m_c^2\over {m_c^2+Q^2}}\alpha_S(m_c^2+Q^2)L(m_c^2+Q^2)
\label{eq:FLPEN1}
\eea
which is short of one power of $L$, though. 
\begin{figure}[t]
\rule{5cm}{0.2mm}\hfill\rule{5cm}{0.2mm}
\vskip 2.5cm
\rule{5cm}{0.2mm}\hfill\rule{5cm}{0.2mm}
\psfig{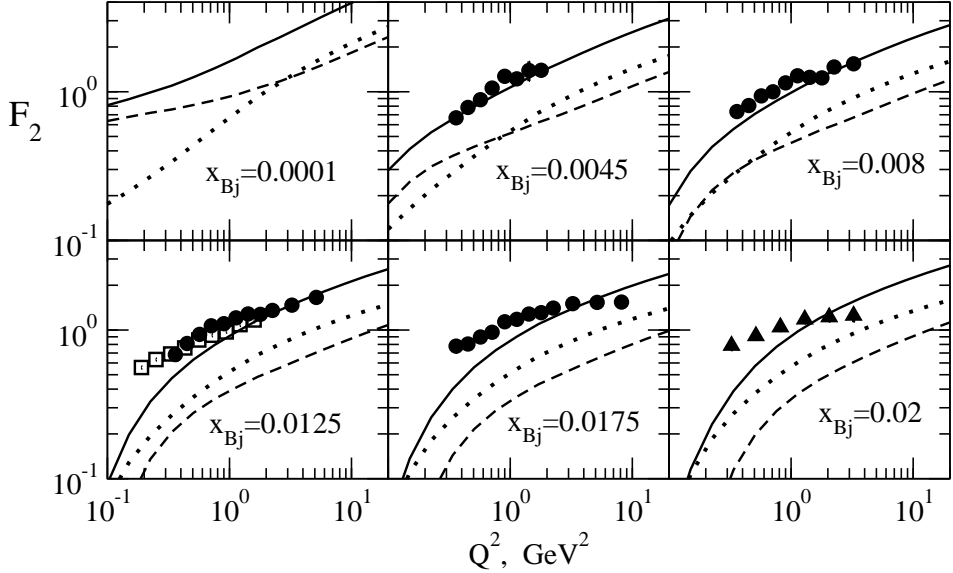}
\caption{The nucleon structure function $F_2$ at smallest available  
$x_{Bj}$ as measured in   $\nu Fe$ CC DIS by the 
CCFR [4]
(circles) and  CDHSW Collaboration [16]
(squares, $x_{Bj}=0.015$). 
Triangles are  the 
CHORUS Collaboration  measurements [17] 
of $F_2$  in 
 $\nu Pb$ CC DIS. 
 Solid curves show the vacuum  exchange 
contribution to $F_2$.
Also shown are the charm-strange (dashed curves) and light 
flavor (dotted curves) components of $F_2$.
\label{fig:fig1}}
\end{figure}  
The rise of $F^{cs}_L(x,Q^2)$ towards small $x$ is generated by interactions
of the higher Fock states,  $c\bar{s}+gluons$.
One can estimate
the leading contribution to $F^{cs}_L$ associated with the Fock state 
 $c\bar{s}+one\,gluon$, 
\bea
\delta F_L^{cs}\sim {N_cC_F\over 4}
{m_c^2\over {m_c^2+Q^2}}{1\over 3!}L^3(m_c^2+Q^2)\eta \,,
\label{eq:DELTAFL}
\eea
where
$\eta=C_A\log\left({x_0/x}\right)$.

The DLLA resummation at $Q^2\lsim m_c^2$
puts  the P-wave component of $F_L^{cs}$ in the form 
\bea
F_L^{cs}\sim {N_cC_F\over 4}{m_c^2\over {m_c^2+Q^2}}{L(m_c^2+Q^2)\eta^{-1}}
I_2(2\sqrt{\xi}),
\label{eq:DGLAP}
\eea
where 
 $\xi= \eta L(m_c^2+Q^2)$ is the DLLA expansion parameter and
 $I_2(z)\simeq \exp(z)/\sqrt{2\pi z}$ is the Bessel function.
Therefore, $F_L^{cs}$ rises rapidly to small $x$. 

Evidently, the perturbative  mechanism of enhancement described above 
does not work  in the 
light quark ($ud$) channel. Besides, Adler's theorem
allows only a slow rise of $F^{ud}_L(x,0)$ to small $x$ \cite{FZCS}, 
$
F^{ud}_L(x,0)\propto (1/x)^{\Delta_{soft}},
$
where $\Delta_{soft}\simeq 0.08$.

At $Q^2\gg m_c^2$ for $\sigma(x,r)
\approx {\pi^2 r^2\over N_c}\alpha_S(r^{-2})G(x,r^{-2})$  
from (\ref{eq:FACTOR}) and (\ref{eq:RHOS1}) 
 it follows that 
$
F^{cs}_L
\sim \alpha_S(Q^2)G(x,Q^2),
$
what  corresponds to the dominance of 
``non-partonic'' configurations with  $z\sim 1/2$ \cite{KolyaNUDIS}. 
 Here  $G(x,k^2)=xg(x,k^2)$ is the 
gluon structure function.

We evaluate $F_L$, $F_T$ and $F_2=F_L+F_T$, 
for the $\nu Fe$ and $\nu Pb$ interactions 
  making use of the  
 approach to nuclear shadowing  developed in \cite{NSZZ}.
The $\log(1/x)$-evolution  
is described by the CD BFKL equation with boundary condition 
at $x_0=0.03$. In Fig. {\ref{fig:fig1}  our results 
(valence-quark contributions are neglected)  are compared with 
experimental data. We conclude that the excitation of charm 
contributes significantly to $F_2$ at $x\lsim 0.01$ and dominates $F_2$
at $x\lsim 0.001$ and $Q^2\lsim m_c^2$.  The agreement with 
data is quite reasonable but it should be taken with some  caution. 
The point is that  the perturbative light-cone 
density of $u\bar d$ states, 
$|\Psi^{ud}|^2\sim r^{-2}$, apparently overestimates the role  of 
short distances at small $Q^2$  and gives the value of  $F_L^{ud}(x,0)$ 
smaller than that required by Adler's theorem  \cite{FZAdler}.
This may lead to underestimation of  $F_2$ in the region of 
moderately  small $x\gsim 0.01$ dominated by the $ud$-current.

\section*{Acknowledgments}
One of the authors (VRZ) is grateful to the organizers for the 
invitation to deliver this talk and support.
The work was supported in part by the Ministero Italiano
dell'Istruzione, dell'Universit\`a e della Ricerca and  by
 the RFBR grant 06-02-16905  and 07-02-00021.

\end{document}